\documentclass{article}
\usepackage{spconf,amsmath,graphicx}
\usepackage{amsfonts} 
\usepackage{subfigure}
\usepackage{color}
\usepackage{cite}
\usepackage{multirow}
\usepackage{threeparttable}
\usepackage{booktabs}
\usepackage{tabularx}
\newcolumntype{C}[1]{>{\centering\arraybackslash}p{#1}}
\newcolumntype{L}[1]{>{\raggedright\arraybackslash}p{#1}}
\newcolumntype{R}[1]{>{\raggedleft\arraybackslash}p{#1}}

\title{Spatial data augmentation with simulated room impulse responses\\for sound event localization and detection}
\name{Yuichiro Koyama$^{\star}$,
\quad Kazuhide Shigemi$^{\dagger *}$,
\quad Masafumi Takahashi$^{\star}$,
\quad Kazuki Shimada$^{\star}$,}

\secondlinename{Naoya Takahashi$^{\star},$
\quad Emiru Tsunoo$^{\star}$,
\quad Shusuke Takahashi$^{\star}$,
\quad Yuki Mitsufuji$^{\star}$}
 
\address{$^{\star}$ Sony Group Corporation, Tokyo, Japan \qquad\qquad
      $^{\dagger}$The University of Tokyo, Japan}

\begin{document}
\ninept
\maketitle
\begin{abstract}
Recording and annotating real sound events for a sound event localization and detection (SELD) task is time consuming, and data augmentation techniques are often favored when the amount of data is limited. However, how to augment the spatial information in a dataset, including unlabeled directional interference events, remains an open research question. Furthermore, directional interference events make it difficult to accurately extract spatial characteristics from target sound events. To address this problem, we propose an impulse response simulation framework (IRS) that augments spatial characteristics using simulated room impulse responses (RIR). RIRs corresponding to a microphone array assumed to be placed in various rooms are accurately simulated, and the source signals of the target sound events are extracted from a mixture. The simulated RIRs are then convolved with the extracted source signals to obtain an augmented multi-channel training dataset. Evaluation results obtained using the TAU-NIGENS Spatial Sound Events 2021 dataset show that the IRS contributes to improving the overall SELD performance. Additionally, we conducted an ablation study to discuss the contribution and need for each component within the IRS.
\end{abstract}
\renewcommand{\thefootnote}{\fnsymbol{footnote}}
\footnote[0]{
$*$Work done during an internship at Sony Group Corporation.}

\section{Introduction}
\label{sec:intro}
Sound event localization and detection~(SELD) involves identifying both the direction of arrival~(DOA) and the type of sound~\cite{adavanne2019sound,cao2019polyphonic,politis2020dataset,nguyen2020sequence,politis2021dataset}. As many combinations of DOA and types of sound are possible, recording and annotating real sound events for a SELD task is time consuming. Therefore, numerous methods have been utilizing data augmentation techniques based on given datasets~\cite{cao2021improved,nguyen2021general,shimada2021accdoa,wang2021four}.

The multi-channel simulation framework (MCS), which convolves extracted covariance matrices with enhanced source signals, enables us to create new combinations by randomly combining extracted spectral information and spatial information~\cite{wang2021four}. Compared with other spatial data augmentations such as the rotation augmentation method~\cite{mazzon2019first}, the framework can change both the directional information and the reverberation pattern of sound events. Furthermore, the MCS does not need clean sources since it extracts enhanced source signals. Experimental results obtained with the TAU-NIGENS Spatial Sound Events 2020 dataset, consisting of only target sound events and diffusive background noise, showed that the MCS helps to improve the SELD performance~\cite{politis2020dataset,wang2021four}.

Directional interference events, which are sound events not included in target sound event classes, can be recorded unintentionally during the recording process because it is generally difficult to perfectly control all sound events generated in a real environment. Contamination by such interference events decreases the number of events that are available as non-overlapped events, i.e., clean events. As a result, extracting spatial information becomes challenging, and this could lead to degradation in the performance of existing data augmentation methods such as the MCS. 
Therefore, we assume that generating the target spatial information with an acoustic simulation could improve the
augmentation quality.  

In this paper, we propose an impulse response simulation framework (IRS) that augments spatial characteristics using simulated room impulse responses (RIR), which is not affected by directional interference events. RIRs corresponding to a microphone array assumed to be placed in various rooms are accurately simulated using image source methods~\cite{Scheibler2018} and spherical-harmonic-domain representation of its frequency response~\cite{moreau2006, Archontis2017}. The source signals of the target sound events are extracted from a mixture on the basis of annotated information and a proposed interference elimination process. The simulated RIRs are then convolved with the extracted source signals to obtain an augmented multi-channel training dataset.
Evaluation results obtained using the TAU-NIGENS Spatial Sound Events 2021 dataset~\cite{wang2021four} show that the IRS contributes to improving the SELD score and the need for each component within the IRS. We also show that the IRS is usable with other typical data augmentation techniques and contributes to achieving state-of-the-art performance.

\section{Related Work}
\label{sec:related_work}
\begin{figure*}[htb]
\centering\subfigure[Workflow of MCS.]{\includegraphics[width=13cm]{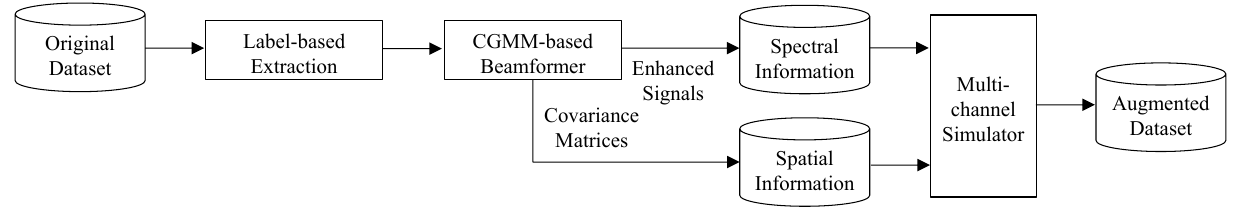}
 \label{fig:mcs}}
\hspace{0mm}
\centering\subfigure[Workflow of IRS. Dotted line represents workflow of MCS with interference elimination block, which is used only for comparison in section~\ref{sec:experiment}.]{\includegraphics[width=17cm]{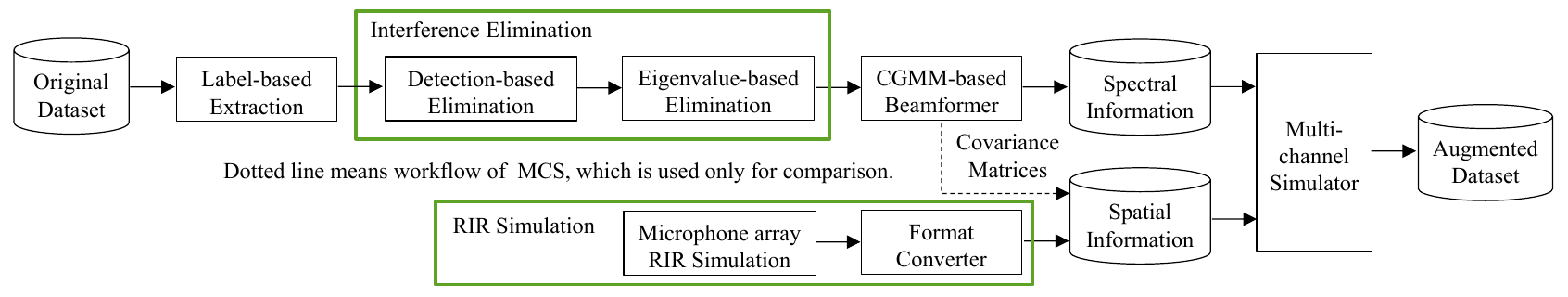}
 \label{fig:irs}}
\hspace{0mm}
\vspace{-2mm}
\caption{Comparison of MCS and IRS. Interference elimination block and RIR simulation block are main differences between MCS and IRS. Interference elimination block is necessary for dealing with dataset including directional interference events.}
\label{fig:workflow}
\vspace{-2mm}
\end{figure*}

\subsection{Supervised approaches for SELD}
The primary problem in the SELD task is how to associate sound event detection (SED) predictions with  DOA predictions (or vice versa) when multiple sound events overlap, which is called the data association problem~\cite{adavanne2019sound}.
To solve this problem, supervised approaches using deep neural network such as the convolutional recurrent neural network (CRNN) are typically used.  

Spectral information, e.g., multi-channel spectrogram, log-mel spectrogram, and spatial information, e.g., GCC-PHAT, inter-channel phase differences (IPDs), are combined and used as input features. These features are fed into several two-dimensional convolutional layers followed by recurrent layers such as a gated recurrent unit~(GRU). For instance, the assigning of a densely connected dilated DenseNet (D3Net)~\cite{takahashi2021densely} to the convolutional layers achieved state-of-the-art performance in previous DCASE challenges~\cite{shimada2021accdoa,shimada2021ensemble}. Finally, the output of the recurrent layers is transformed by fully-connected layers into an output representation. While two-branch representation, e.g., SELDnet~\cite{adavanne2019sound}, uses two branches for two targets, an SED target and a DOA target, activity-coupled Cartesian DOA~(ACCDOA) representation assigns a sound event activity to the length of a corresponding Cartesian DOA vector~\cite{shimada2021accdoa}. The ACCDOA representation enables us to handle a SELD task as a single task with a single network.

\subsection{Data augmentation}

Some data augmentation techniques have been widely used for the SELD task. EMDA~\cite{Takahashi16,Takahashi2017AENet} mixes audio events with random amplitudes, delays, and the modulation of frequency characteristics, i.e., equalization. The rotation augmentation method~\cite{mazzon2019first} rotates an observed signal represented in the first-order Ambisonics (FOA) format and enables us to increase the number of DOA labels without losing the physical relationships between steering vectors and observations. SpecAugment~\cite{park2019specaugment}, which was originally proposed for the speech recognition task and recently also showed efficacy on the SELD task~\cite{shimada2021accdoa}, applies frequency masking to input features.

In addition to these augmentation methods, the MCS is also effective for creating new combinations of spectral information and spatial information~\cite{wang2021four}. Fig.~\ref{fig:mcs} describes the workflow of the MCS. First, non-overlapping and static, i.e., not moving, events are extracted from an original dataset on the basis of label information. Then, the complex Gaussian mixture model~(CGMM)-based beamformer~\cite{higuchi2016} block takes the extracted events and outputs enhanced signals and covariance matrices. 
The covariance matrices are calculated with the spectrograms enhanced by CGMM-based masking.
The enhanced signals are stored as spectral information, and the covariance matrices are stored as spatial information. They are independently and randomly sampled to create new combinations of spectral information and spatial information. The multi-channel simulator block finally takes them and simulates multi-channel input signals. 

The MCS actually contributed to improving the performance on the TAU-NIGENS Spatial Sound Events 2020 dataset~\cite{politis2020dataset,wang2021four}, which does not include directional interference events. However, it assumes no directional interference events and highly relies on non-overlapping events (i.e., clean events) in the original dataset. Therefore, using the MCS for datasets including directional interference events such as the TAU-NIGENS Spatial Sound Events 2021 dataset could lead to performance degradation.

\vspace{-2.9mm}
\section{Proposed method}
\label{sec:proposed_method}

Inspired by the MCS, we propose the IRS, which augments spatial characteristics using RIRs. Fig.~\ref{fig:irs} describes the workflow of the IRS. The interference elimination block and RIR simulation block are the main differences from the MCS. The interference elimination block is necessary for dealing with datasets including directional interference events such as the TAU-NIGENS Spatial Sound Events 2021 dataset~\cite{politis2021dataset}. Also, we assume that the covariance matrix of an event extracted from the original dataset is still not clean enough for use as spatial information even if the interference elimination block is applied. Therefore, we simulate RIRs, which are used for spatial information.

\subsection{Interference elimination}

The interference elimination block is composed of two blocks: the detection-based elimination block and eigenvalue-based elimination block. The motivation for using two types of elimination block is to eliminate events with different ranges of signal-to-interference ratio (SIR).

First, the detection-based elimination block uses a model pre-trained without the IRS, whose performance is shown as ID~0 in Table~\ref{tab:result_ablation} . The non-overlapping static events extracted by the label-based extraction block are processed with the pre-trained model. Events that the pre-trained model cannot detect are regarded as events overlapped with interference events, which are eliminated.

Then, the eigenvalue-based elimination block first applies a short-time Fourier transform~(STFT) to the events extracted by the detection-based elimination block and computes a spatial covariance matrix from the obtained spectrogram. Its eigenvalues are then computed and normalized such that the maximum eigenvalue becomes 1.
These eigenvalues can be considered to reflect how many sound sources the observation signals include.
Let $\gamma_c (c=1,\dots,C)$ be the normalized eigenvalues, where $C$ is the number of input channels.
We introduce two types of thresholds, $\alpha$ and $\beta$. 
We define the frequency bin where the number of eigenvalues that satisfy $\gamma_c > \alpha$ is more than 1 as an \emph{overlapped bin}. Then, focusing on the limited range from the minimum frequency $f_{\text{min}}$ to maximum frequency $f_{\text{max}}$, let $K_{\text{focus}}$ be the number of total focused bins and $K_{\text{overlap}}$ be the number of \emph{overlapped bins}. Extracted events that satisfy $(K_{\text{overlap}}/K_{\text{focus}}) > \beta$ are regarded as events overlapped by interference events, which are also eliminated.

\begin{figure}[tb]
 \centering
 \includegraphics[width=8.0cm]{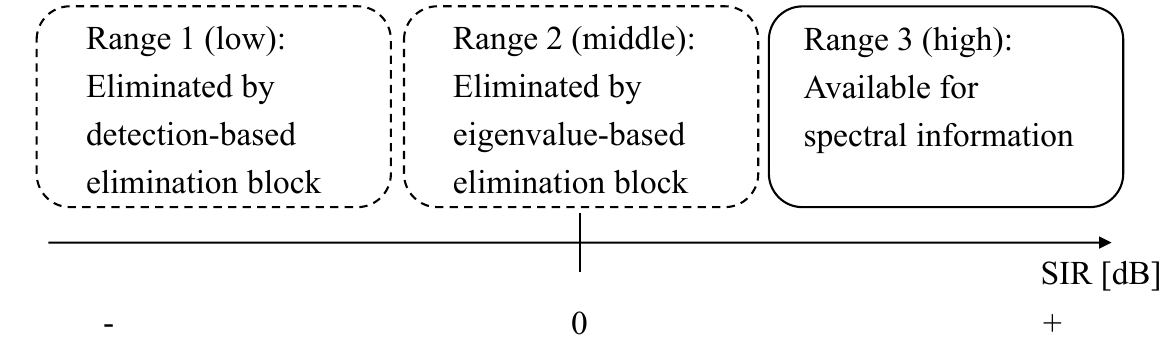}
 \vspace{-1mm}
\caption{Assumption in terms of SIR. We propose using events from range 3 for spectral information.}
\label{fig:sir}
\vspace{-4mm}
\end{figure}

Fig.~\ref{fig:sir} illustrates our assumption regarding the relationship between the interference elimination block and SIR, where we divide the SIR range into three levels: ``range~1 (low)," ``range~2 (middle)," and ``range~3 (high)." We assume that the events in range~1 shown in Fig.~\ref{fig:sir} can be eliminated by the detection-based elimination block because the events here are heavily overlapped by interference events; thus, the pre-trained model cannot detect such events. The events in range~2 are expected to be eliminated by the eigenvalue-based elimination block because the power of a target sound event and that of an interference event are almost the same and  $K_{\text{overlap}}/K_{\text{focus}}$ tends to be higher. The events in range~3 are fed into the CGMM-based beamformer block, and the SIRs are improved by the beamformer. As a result, the enhanced signals are clean enough to be used as spectral information.

\subsection{RIR simulation}
The RIR simulation block is composed of two blocks, the
microphone-array RIR simulation block and format converter block, 
by which spatial aliasing caused by the discrete sampling of a sound field can be simulated to reflect the actual recording condition.

In the microphone-array RIR simulation block, 
an image source method~\cite{Scheibler2018} is utilized to define the positions for all image sources depending on given reverberant conditions. 
RIRs are obtained as linear combinations of the frequency responses $H$ for all image sources.
We assume that the array can be regarded as a rigid spherical array~\cite{Politis2016, MitsufujiTKS21}. 
By considering the waves from all image sources as plane waves, the frequency response of $h$-th microphone with a wave number of $k$ on a rigid baffle of radius $R$ for $l$-th image source is obtained as
\begin{equation}
H_{hl}(k, \psi_{hl}) =  g_{hl} e^{-\mathrm{i} k d_{hl}}\sum_{n=0}^{\infty} \mathrm{i}^n ( 2n+1) b_n\left(k R\right) P_n\left(\cos\psi_{hl}\right),
\label{eq:rigid_array_response}
\end{equation}
where $\psi_{hl}$ denotes the angle between the DOA of the $l$-th plane wave and the orientation of the $h$-th microphone, $P_{n}(\cdot)$ denotes the Legendre polynomial~\cite{Williams1999, Politis2016}, 
$g_{hl}$ and $d_{hl}$ denote attenuation factor and delay time respectively, both of which are caused by the number of reflections and distance. The imaginary unit is denoted by $\mathrm{i}$.
The $b_n$ is a radial function for a rigid baffle array written as
\begin{equation}
 b_n(k R) = \frac{\mathrm{i}}{(k R)^2 h_n^{(1)'}(k R)},\label{eq:bn}
\end{equation}
where $h^{(1)'}_n(\cdot)$ denotes the derivative of the $n$-th-order spherical Hankel function of the first kind. 
Computing the linear combinations of $H_{hl}(k, \psi_{hl})$ in terms of all image sources, the simulated RIR of $h$-th microphone can be obtained as
\begin{equation}
    x_h(k) = \sum_{l} H_{hl}(k, \psi_{hl}) .\label{eq:rir}
\end{equation}

Then, in the format converter block, 
the simulated RIRs are converted to the intended format, e.g., higher-order Ambisonics~(HOA). 
The $n$-th-order and $m$-th-degree spherical harmonic function is defined with the angle $\Omega=\{\theta, \phi\}$~\cite{Williams1999} as

\begin{equation}
Y_{nm}(\Omega) \equiv \sqrt{\frac{2n+1}{4 \pi}\frac{(n-m)!}{(n+m)!}} P_{nm}(\cos \theta) e^{\mathrm{i} m \phi},
\label{eq:Ynm}
\end{equation}
where $\theta$, $\phi$, and $P_{nm}(\cdot)$ denote the elevation, azimuth, and associated Legendre function, respectively. 
The spherical-harmonic representation of the RIR can be computed by using the following encoding process~\cite{moreau2006, Archontis2017}:
\begin{equation}
 \mathbf{a}(k) = \mathbf{B}(k)^{-1} \mathbf{Y}^{\dag} \mathbf{x}(k),
 \label{eq:mtx_SHD_conv}
\end{equation}
with
\begin{equation}
\label{eq:bessel_matrix}
\mathbf{B}(k) = \left(
 \begin{array}{cccccc}
 b_0 & 0 & 0 & 0 & \ldots & 0 \\
 0 & b_1 & 0 & 0 & \ldots & 0 \\
 0 & 0 & b_1 & 0 & \ldots & 0 \\
 0 & 0 & 0 & b_1 & \ldots & 0 \\
 \vdots & \vdots & \vdots & \vdots & \ddots & \vdots \\
 0 & 0 & 0 & 0 & \ldots & b_{N}
 \end{array}
 \right), 
\end{equation}

\begin{equation}
\label{eq:SHT_matrix}
\mathbf{Y} = 
 \begin{bmatrix} \mathbf{y}(\Omega_{1}) & \mathbf{y}(\Omega_{2}) & \cdots & \mathbf{y}(\Omega_{M})
 \end{bmatrix}^T,
\end{equation}
where $\mathbf{y}(\Omega) \in \mathbb{C}^{(N+1)^2}$ is a column vector containing $Y_{nm}(\Omega)$, $N$ is the spherical-harmonic order, e.g., $N=1$ corresponds to FOA, $M$ is the number of microphones, $\mathbf{x}(k) \in \mathbb{C}^{M}$ is a column vector containing $x_h(k)$, 
and $(\cdot)^{\dag}$ represents the Moore--Penrose pseudoinverse.


\section{Experiment}
\label{sec:experiment}

\subsection{Experimental settings}

We evaluated our approach on the development set of the TAU-NIGENS Spatial Sound Events 2021 dataset using the suggested setup~\cite{politis2021dataset}. The dataset was composed of 6 folds, which contained 600 one-minute Eigenmike recordings with the FOA format: 400 for training (folds 1 to 4), 100 for validation (fold 5), and 100 for testing (fold 6). The sound event samples were from the NIGENS general sound events database~\cite{trowitzsch2019nigens}, which consists of 12 event classes such as footsteps and barking dog. Each event had an equal probability of being either static or moving. The signal-to-noise ratios ranged from 6 dB to 30 dB. The sampling frequency was 24 kHz.

We prepared two types of CRNNs: the conventional CRNN used in~\cite{cao2019polyphonic} and RD3Net~\cite{takahashi2021densely}. 
Both networks have four convolutional blocks followed by a GRU. They take frame-wise multi-channel amplitude spectrograms and inter-channel phase differences~(IPDs) as frame-wise features and output frame-wise ACCDOA. The STFT was applied with a configuration having a 20-ms frame length and 10-ms frame hop.

\begin{table*}[ht]
    \caption{Experimental results with CRNN-based architecture on TAU-NIGENS Spatial Sound Events 2021 dataset. D-based and E-based represents detection-based elimination and eigenvalue-based elimination respectively. CM stands for covariance matrix.}
    \label{tab:result_ablation}
    \centering
    \scalebox{0.90}{
    \begin{tabular}{l|cccc|c|ccccc}
        \toprule
        \multirow{2}{*}{ID} & \multirow{2}{*}{IRS} & \multicolumn{2}{c}{Interference elimination} & \multicolumn{1}{c|}{Spatial} & \multicolumn{1}{c|}{Other} & \multicolumn{5}{c}{Metrics in test set} \\
         & & D-based & E-based & information & augmentation & $\text{ER}_{20} \downarrow$ & $\text{F}_{20} \uparrow$ & $\text{LE}_{\text{CD}} \downarrow$ & $\text{LR}_{\text{CD}} \uparrow$ & $\text{SELD}_{\text{score}}\downarrow$ \\ 
        \midrule
        0 & None & & & & None & 0.68 & 41.2 & 18.1 & 40.8 & 0.489 \\
        \midrule
        1 & \checkmark(MCS) & None & None & CM & None &
        0.66 & 43.6 & 19.0 & 44.4 & 0.471\\ 
        2 & \checkmark(MCS) & \checkmark & \checkmark & CM & None & 0.64 & 45.6 & 18.0 & 46.1 & 0.457  \\ 
        3 & \checkmark & None & None & RIR & None & 0.64 & 46.6 & 17.2 & 44.9 & 0.455 \\ 
        4 & \checkmark & \checkmark & None & RIR & None & 0.64 & 46.4 & 18.3 & 45.7 & 0.455 \\ 
        5 & \checkmark & None & \checkmark & RIR & None & 0.63 & 47.7 & 17.7 & 47.1 & 0.445 \\ 
        6 & \checkmark & \checkmark & \checkmark & RIR & None & 0.62 & 49.2 & 16.8 & 48.1 & \textbf{0.436} \\       
        \midrule
        7 & None & & & & \checkmark & 0.56 & 55.0 & 18.4 & 59.5 & 0.380  \\ 
        \midrule
        8 & \checkmark(MCS) & None & None & CM & \checkmark & 0.58 & 53.6 & 18.0 & 59.4 & 0.386 \\ 
        9 & \checkmark(MCS) & \checkmark & \checkmark & CM & \checkmark & 0.57 & 54.9 & 18.1 & 59.4 & 0.381  \\ 
        10 & \checkmark & None & None & RIR & \checkmark & 0.56 & 55.2 & 18.0 & 60.3 & 0.377  \\ 
        11 & \checkmark & \checkmark & None & RIR & \checkmark & 0.56 & 56.1 & 17.3 & 60.2 & 0.372 \\ 
        12 & \checkmark & None & \checkmark & RIR & \checkmark & 0.54 & 56.0 & 16.2 & 57.0	& 0.376 \\ 
        13 & \checkmark & \checkmark & \checkmark & RIR & \checkmark & 0.54 & 57.8 & 17.0 & 61.5 & \textbf{0.360}  \\       
        \bottomrule
    \end{tabular}
    }
\vspace{-4mm}
\end{table*}
The IRS was applied to the training data by adding new training folds created by the IRS. 
In this experiment, we added 2~folds, which increased the amount of training data by 50\%. 
The thresholds for our interference elimination block, $\alpha$ and $\beta$, were 0.3 and 0.4 respectively. The $f_{\text{min}}$ and $f_{\text{max}}$ were 100~Hz and 4~kHz, respectively. A minimum variance distortionless response (MVDR) beamformer~\cite{souden2009optimal} was applied to the CGMM-based beamformer block. 
In the RIR simulation block, the reverberation time~(RT60) was randomly set to be within 100 to 500~ms.
We only simulated static impulse responses. 
The SN3D normalization scheme of Ambisonics was used in the format converter block~\cite{politis2020dataset}.
For comparison, the MCS was also applied by using covariance matrices for the spatial information instead of RIRs. In addition, we investigated the efficiency of combining the IRS and other augmentation techniques, that is, EMDA, rotation augmentation methods, and a multi-channel version of SpecAugment, whose efficacy were already shown in \cite{shimada2021ensemble}. These augmentation techniques were applied on-the-fly~\cite{erdogan2018investigations}.

The frame length of the network input during training was 128 frames except for the experiment to show the performance of RD3Net. The batch size for the training was 32. The learning rate was linearly increased from 0.0 to 0.001 with 50,000 iterations~\cite{goyal2017accurate}. After the warm-up, the learning rate was decreased by 10\% if the SELD score of the validation did not improve in 40,000 consecutive iterations. We used the Adam optimizer with a weight decay of~$10^{-6}$. We validated and saved model weights every 10,000 iterations up to 400,000 iterations. Finally, we averaged the model weights from the last 5 models as in \cite{karita2019comparative}.

Four metrics were used for the evaluation~\cite{mesaros2019joint}: $\text{LE}_\text{CD}$, $\text{LR}_\text{CD}$, $\text{ER}_{20^{\circ}}$, and $\text{F}_{20^{\circ}}$. $\text{LE}_\text{CD}$ is a localization error that indicates the average angular distance between predictions and references of the same class. $\text{LR}_\text{CD}$ is a simple localization recall metric that expresses the true positive rate of how many of these localization predictions are correctly detected in a class out of the total number of class instances. $\text{ER}_{20^{\circ}}$ and $\text{F}_{20^{\circ}}$ are the location-dependent error rate and F-score, where predictions are considered as true positives only when the distance from the reference is less than $20^{\circ}$. To evaluate the overall performance, we adopted $\text{SELD}_{\text{score}}$, which is defined as
\begin{equation}
\text{SELD}_{\text{score}} = [\text{ER}_{20^{\circ}} + (1-\text{F}_{20^{\circ}}) + \text{LE}_\text{CD}/\pi + (1-\text{LR}_\text{CD})]/4.
\label{eq:seld}
\end{equation}

\begin{table}
    \caption{Performance of RD3Net}
  \label{tab:result_rd3}
  \centering
 \scalebox{0.90}{
 \begin{tabular}{c|ccccc}
    \toprule

        \multirow{2}{*}{IRS} & 
        \multicolumn{5}{c}{Metrics for test set}  
        
        \\ 
         & $\text{ER}_{20} \downarrow$ & $\text{F}_{20} \uparrow$ & $\text{LE}_{\text{CD}} \downarrow$ & $\text{LR}_{\text{CD}} \uparrow$ & $\text{SELD}_{\text{score}}\downarrow$  \\ 
        \midrule
        None~\cite{shimada2021ensemble} &  
        0.48 &
        64.1 &
        13.2 &
        63.2 &
        0.321 \\ 
        \checkmark & 
        0.49 &
        65.0 &
        14.0 &
        70.7 &
        \textbf{0.302}\\ 
        \bottomrule
    \end{tabular}
}
\vspace{-3mm}
\end{table}

\subsection{Results}

Table~\ref{tab:result_ablation} shows the experimental results for the IRS using CRNN. ID~0 represents the training method without any data augmentation techniques, and ID~7 represents that with other data augmentation techniques (i.e., EMDA, rotation augmentation method, multi-channel version of SpecAugment). IDs~1 and 8 are equivalent to the MCS, and IDs~2 and 9 are the MCS with the interference elimination block.

The $\text{SELD}_{\text{score}}$ of ID 0, 3, 4, 5, and 6 shows that the IRS consistently improved the $\text{SELD}_{\text{score}}$ and that using both the eigenvalue-based elimination block and detection-based elimination block was the most effective. In this experiment, the number of events extracted by the label-based extraction was 1243, the number of events eliminated by the eigenvalue-based elimination block was 244, and the number of events eliminated by the detection-based elimination block was 16. This result suggests that using the events from range 1 in Fig.~\ref{fig:sir} for the spectral information degrades the performance even if the number of events is relatively small. This is because such events have an extremely low SIR as shown in Fig.~\ref{fig:sir}. Comparing IDs~1, 2, 3, and 6 (i.e., comparing the IRS with the MCS), it is shown that the RIR was more suitable for the spatial information than the covariance matrix calculated from the extracted events. This is because the events from range 3 in Fig.~\ref{fig:sir} still had interference events with low energy, which disturbed the spatial information of the target sound events. The results for IDs~7 to 13 show the same tendency as the results for IDs~0 to 6, which suggests that the IRS can be used with other augmentation techniques.

Table~\ref{tab:result_rd3} shows the performance of RD3Net. We trained RD3Net with other augmentation techniques and 1024 input frames.
It is shown that the IRS contributed to further improving the performance of the state-of-the-art method (i.e., RD3Net). 

\vspace{-1mm}

\section{Conclusion}
We proposed the impulse response simulation framework (IRS), which augments spatial characteristics using simulated room impulse responses. Experimental results obtained using the TAU-NIGENS Spatial Sound Events 2021 dataset indicated that each block in the IRS contributed to improving the SELD performance. In addition, it was shown that combining the IRS with other typical augmentation techniques lead to further improvement. Finally, RD3Net trained with the IRS achieved state-of-the-art performance on the SELD task. 
In future work, we will explore a new framework that does not depend on the label-based extraction block, which will help us to create new combinations of spectral and spatial information without supervised annotations.

\vfill\pagebreak

\bibliographystyle{IEEEtran}
\bibliography{strings,refs}

\end{document}